# Deterministic Electrical Control of Single Magnetic Bubbles in Nanostructured Cells

*Jialiang Jiang, Yaodong Wu*, Lingyao Kong, Yongsen Zhang, Sheng Qiu, Huanhuan Zhang, Yihao Wang, Junbo Li, Yimin Xiong, Shouguo Wang, Mingliang Tian, Haifeng Du*, and Jin Tang**

Jialiang Jiang, Lingyao Kong, Huanhuan Zhang, Yimin Xiong, Mingliang Tian, Jin Tang

School of Physics and Optoelectronic Engineering, Anhui University, Hefei, 230601, China

Email: wuyaodong@hfnu.edu.cn; duhf@hmfl.ac.cn; jintang@ahu.edu.cn

Yaodong Wu

School of Physics and Materials Engineering, Hefei Normal University, Hefei 230601, China

Yaodong Wu, Sheng Qiu, Yihao Wang, Mingliang Tian, Haifeng Du, Jin Tang

Anhui Provincial Key Laboratory of Low-Energy Quantum Materials and Devices, High Magnetic Field Laboratory, HFIPS, Chinese Academy of Sciences, Hefei 230031, China

Yongsen Zhang, Shouguo Wang

Anhui Provincial Key Laboratory of Magnetic Functional Materials and Devices, School of Materials Science and Engineering, Anhui University, Hefei 230601, China

**Abstract**

Localized particle-like spin textures have been found to exhibit emergent electromagnetic properties, which hold promise for the development of intriguing spintronic devices. Among these textures, magnetic bubbles represent localized spin configurations that could serve as data bits. However, the precise methods for their electrical manipulation remain uncertain. Here, we demonstrate the deterministic electrical manipulations and detections of single magnetic bubbles in kagome-latticed $Fe_3Sn_2$ magnetic nanostructured cells. The current-induced dynamics of magnetic bubbles were explored using nanosecond pulsed currents. We show single pulsed currents with low and high densities can be applied for the creation and deletion of a single bubble, respectively. The mutual writing-deleting operations on single bubbles are attributed to the thermal heating and non-thermal spin-transfer torque effects in combination with micromagnetic simulations. We also realized the in-situ detection of a single bubble using the anisotropic magnetoresistance effect through a standard four-probe method. Our results could propel the development of bubble-based spintronic devices.

# 1. Introduction

Particle-like localized magnetic textures, such as skyrmions, serve as natural data bits suitable for application in spintronic devices due to their high speed, high density, and low energy consumption.[1-9] Recent research has demonstrated that skyrmions exhibit a strong coupling with electrical current, resulting in ultralow depinning critical currents,[1, 10] topological Hall signals,[11, 12] and the skyrmion Hall effect.[13, 14] This rich electrodynamics enables the electrical creation, deletion, driving, and detection of skyrmions.[15-20] While other magnetic spin textures, such as magnetic antiskyrmions,[21, 22] skyrmion bundles,[23-25] bobbers,[26] bubbles,[27] and hopfions,[28] have also garnered significant attention for their unique physical properties.

Magnetic bubbles are cylindrical domains that were discovered in centrosymmetric uniaxial ferromagnets in the last century. When stimulated by oscillating in-plane fields, magnetic bubbles can be steadily driven in "T-I" shaped geometries due to magnetic dipole-dipole interaction.[29, 30] They are considered potential candidates for data bits in magnetic bubble memory, where data operations are carried out through bubble motions.[31] In 1977, a 68 KB prototype of a bubble-based memory device was introduced, marking the start of commercial products.[32] However, this technology later lost out to hard disk drives (HDD) due to limitations in memory density and speed caused by the magnetic field method. Recently, the growing field of spintronics has reignited interest in magnetic spin textures, including magnetic bubbles.

In centrosymmetric uniaxial magnets, magnetic bubbles are typically categorized into two types based on the spin arrangement of domain walls.[27] Type-I bubbles, characterized by a closed cylinder domain wall, possess an integer topological charge as that of skyrmions and are commonly referred to as skyrmion bubbles or dipolar skyrmions.[33-35] They are principally stabilized by the magnetic dipole-dipole interaction. On the other hand, type-II bubbles consist of two arch-shaped domain walls with a pair of Bloch lines between them, having a trivial topology.[36] The magnetizations of domain walls in type-II bubbles align with the in-plane magnetic field component, suggesting that a tilted magnetic field aids in stabilizing these bubbles. Another similar spin configuration is called a target bubble, which is composed of multiple concentric skyrmion bubbles.[37] Particularly, the skyrmionium configuration emerges when two nested skyrmion bubbles form a target texture exhibiting zero topological charge ($Q = 0$). The recent discovery of magnetic skyrmions in experiments showcasing abundant electromagnetic dynamics has reignited interest in magnetic bubbles. Additionally, magnetic bubbles can have nanometric dimensions (approximately 50-70 nm) when compared to chiral magnetic skyrmions at room temperature.[33, 38] The two types of bubbles display distinct responses to current-induced torques, with type-I skyrmion bubbles exhibiting steady movement[39] and type-II trivial bubbles being more prone to pinning.[40] The convergence of nanometer-sized magnetic bubbles and the current advancements in spintronics present new opportunities for the development of magnetic bubble-based electronic devices. Recent research indicates that mutual transformations between skyrmion bubbles and trivial bubbles can be achieved by adjusting the current density.[40] However, despite advancements in magnetic bubble-based electronic devices, the deterministic electrical control of a single data bit, encompassing the creation, erasure, and detection of magnetic bubbles, remains a challenge.

In this study, we demonstrate the deterministic electrical control of a single magnetic bubble in confined $Fe_3Sn_2$ nanostructured cells through creation, deletion, and detection processes at room temperature. By applying alternating low and high pulsed currents, we achieve reliable writing and deleting operations of a single magnetic bubble. The results of micromagnetic simulations suggest that spin transfer torques from pulsed currents and magnetic hysteresis are key factors in the electrical control mechanisms. Additionally, we successfully detect a single magnetic bubble using the standard four-probe method based on the anisotropic magnetoresistance (AMR) effect. Our findings show that a single magnetic bubble results in a magnetoresistivity change of approximately –22 nΩ cm compared to the ferromagnetic (FM) state. These results pave the way for further exploration of bubble-based devices and provide an effective approach for the deterministic electrical control of magnetic spin textures in confined nanostructures.

## 2. Results and Discussions

### 2.1. Magnetic Field-Driven Evolution of $Fe_3Sn_2$ Nanostructured Cell

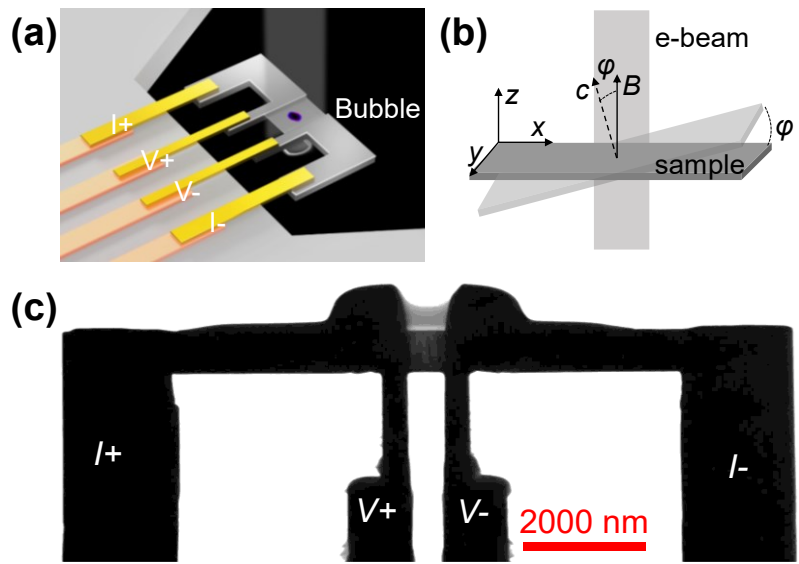

**Figure 1**. Schematic images of the nanostructured cell sample. a) Schematic diagram of the $Fe_3Sn_2$ nanostructured cell sample with a (001) plane. A single magnetic bubble is obtained in the center of the cell. b) Schematic diagram of the incident electron beam interacting with the titled sample plane at an angle $\varphi$. c) TEM images of the sample with four platinum probes.

$Fe_3Sn_2$ is a uniaxial centrosymmetric magnet featuring a kagome lattice composed of Sn and Fe-Sn layers stacked along the $c$-axis.[27, 37, 41-46] It is a bubble-host material with a sufficiently high Curie temperature ($T_c$) reaching ≈ 640 K, which supports its future implementation in spintronic devices.[35] Through the standard focused ion beam (FIB) lift-out method, nanostructured $Fe_3Sn_2$ cells measuring approximately 650 × 550 × 165 nm³ were prepared, incorporating four platinum electrical probes (comprising two current probes and two voltage probes). Notably, the presence of a single magnetic bubble at room temperature was confirmed (**Figure 1** and Supplemental Figure S1). The magnetic field-driven evolutions of magnetic structures in the nanostructured cells are investigated via Lorentz transmission electron microscopy (Lorentz TEM),[47] as displayed in Figure 2.

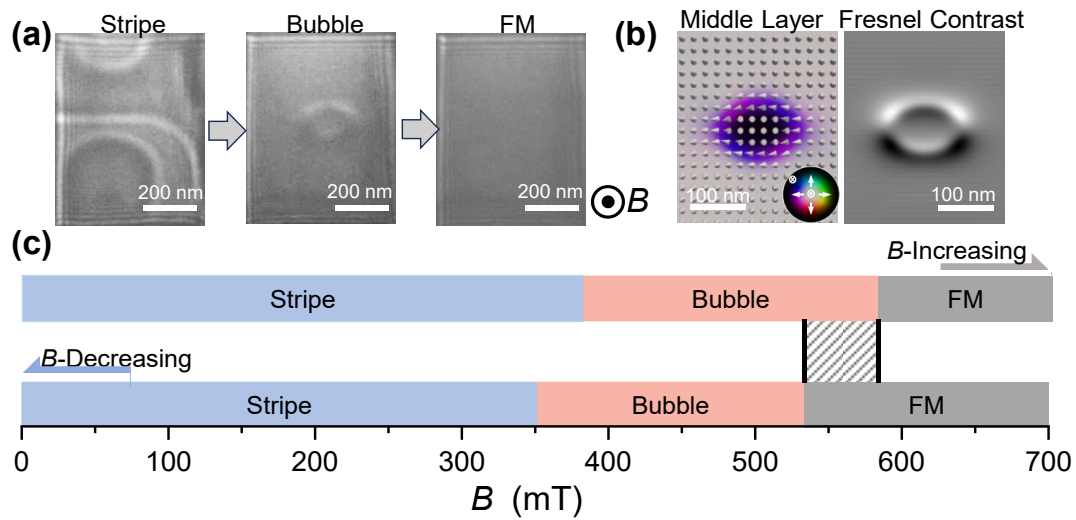

**Figure. 2** Magnetic field-driven evolutions of the nanostructured $Fe_3Sn_2$ cells. a) Three representative defocused Fresnel images recorded by Lorentz TEM in increasing and decreasing field processes, display stripe domains, bubble, and FM respectively. The defocused distance is –750 μm. b) Simulated single magnetic bubble in the middle layer of the cell and its corresponding Fresnel contrast. The color wheel indicates the orientation of magnetization. Dark and white contrasts represent out-of-plane down and up magnetizations, respectively. c) Magnetic phase diagram of field-driven magnetic evolution of the cells. An obvious hysteresis region is marked by the shadow region, where both the bubble and FM state are stable in the field range of 543-582 mT.

To stabilize magnetic bubbles in the nanostructured sample, magnetic field $B$ needs to deviate from the normal direction of the sample by about 3° ($\varphi$), creating sufficient in-plane magnetic field component (Figure 1b). At zero magnetic field, the sample exhibits curving magnetic domains. As the field increases, the magnetic domains shrink at first, eventually transforming into a single magnetic bubble at $B$ = 377 mT. Subsequently, as the magnetic field further increases, the bubble gradually diminishes and transitions into a ferromagnetic (FM) state at $B$ = 582 mT (**Figures 2**a and 2c). In the subsequent magnetic field decreasing process, the FM state remains stable at first, then turns to a magnetic bubble at $B$ = 543 mT, and at last extends to stripe domains at $B$ = 353 mT. Notably, a clear hysteresis effect is present in the magnetic field range of approximately 543-582 mT, highlighted by the shaded region in Figure 2, indicating the stable coexistence of the magnetic bubble and FM state. This situation suggests that the energy of the magnetic bubble and FM state can be considered equivalent within this magnetic field range, enabling a potential electrically controlled transition between these two states. Furthermore, utilizing micromagnetic simulations, we investigated the magnetic evolutions of the nanostructured cells, revealing a consistent magnetic hysteresis phenomenon (Supplemental Figure S2).

## 2.2 Writing and Deleting a Single Bubble

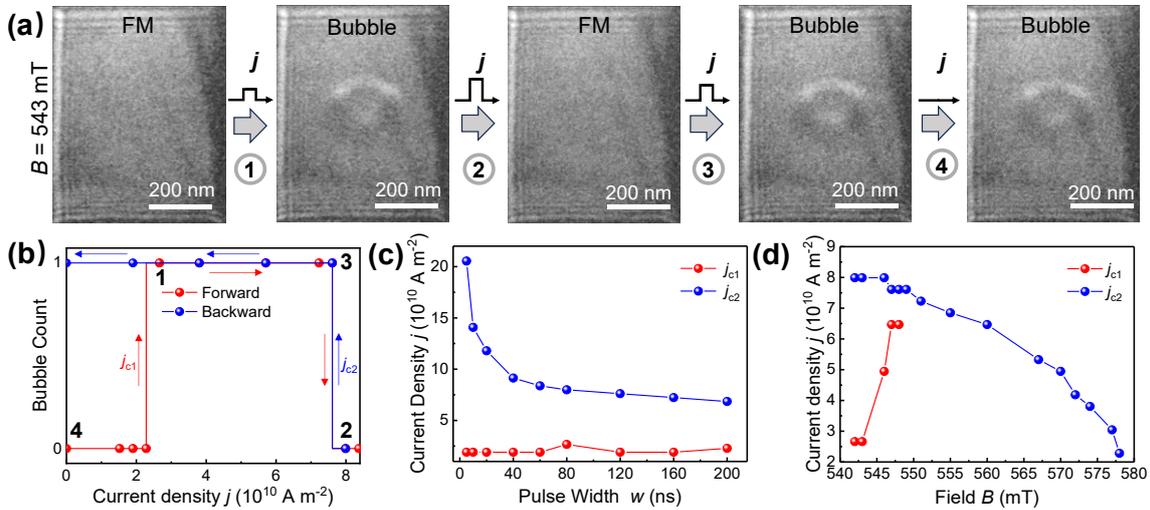

**Figure 3.** Writing and deleting a single bubble by pulsed currents. a) Defocused images of different magnetic states in the nanostructured cell after applying pulsed currents. The defocused distance is –750 μm. Serial numbers 1-4 are corresponding to distinct current densities $j$ = 2.66, 7.99, 7.61, and 0 × $10^{10}$ A m$^{-2}$ in (b). $B$ = 543 mT. b) The relationship between the bubble count and the current density $j$. $j_{c1}$ (red points) and $j_{c2}$ (blue points) are defined as critical current densities for the transition from the FM state to a bubble and vice versa, respectively. $B$ = 543 mT. c) Dependence of critical current densities $j_{c1}$ and $j_{c2}$ on pulse width $w$ at a fixed magnetic field, $B$ = 543 mT. d) Dependence of critical current densities $j_{c1}$ and $j_{c2}$ on magnetic field $B$ at a fixed pulse width, $w$ = 80 ns.

Choosing the FM state as the initial state, we explored the evolutions of the magnetic states under varied pulsed currents with a pulse width of 80 ns, at $B$ = 543 mT (**Figures 3**a and 3b). Initially, at low current densities, the sample maintains the FM state until it reaches $2.66 \times 10^{10}$ A m$^{-2}$, then a magnetic bubble appears due to the STT effect induced by the electrical current. Upon further current density increments, the magnetic bubble persists until transitioning back to the FM state at $7.99 \times 10^{10}$ A m$^{-2}$. The FM state switches back to a magnetic bubble at a current density of $7.61 \times 10^{10}$ A m$^{-2}$, with the bubble state persisting even as the current density returns to zero. The distinct behavior exhibited during the transformation of the magnetic state concerning various pulsed currents allows for the unambiguous identification of two critical current densities, $j_{c1}$, and $j_{c2}$ (Figure 3b). Here, $j_{c1}$ is the lower critical current density marking the transition from the FM state to the bubble state, while $j_{c2}$ is the higher critical current density indicating the transition from the bubble state to the FM state. Thus, the creation and deletion of a single magnetic bubble can be controlled by applying pulsed currents to the nanostructured cells, which includes two important transition processes—from the FM state to the magnetic bubble state and vice versa. Next, combined with the micromagnetic simulations, the underlying physical mechanisms governing these transition processes will be discussed.

For the first transition process from the FM state to the magnetic bubble state, the STT effect induced by the electrical pulsed current plays an important role, and we performed micromagnetic simulations to explain it. The fundamental magnetic parameters in simulations are defined based on Fe$_3$Sn$_2$ material at room temperature. As illustrated in **Figure 4**, the three-dimensional spin textures of the 165-nm thick nanostructured cells under pulsed current are simulated. At the field $B$ = 365 mT, a FM state is obtained at $t$ = 0 ns with a uniform magnetization in the middle layer of the sample (Figure 4a). It is worth knowing that the FM state at that field is not fully uniform, as both the top and bottom surfaces appear vortex-like spin textures, also called surface spin twists (Supplemental Figure S3). Prior research has confirmed the existence of these surface spin twists in 3D skyrmion bubbles within $Fe_3Sn_2$ nanostructures, attributed to the magnetic dipole-dipole interaction (DDI).[27] Therefore, the depth-modulated surface spin twists of the FM state are also influenced by the DDI, as corroborated by an investigation on the anisotropic magnetoresistance of $Fe_3Sn_2$.[17] Upon the application of an electrical current with a density of $j$ = 50 × $10^{10}$ A m$^{-2}$, the surface spin twists undergo distortion and warping. At $t$ = 2 ns, non-uniform magnetizations, deviating from the magnetic field alignment, emerge in the central region of the nanostructured cell, decreasing the dipolar energy of the system (Figures 4a and 4b). These non-aligned magnetizations progressively propagate through all layers of the nanostructure, transforming into a distorted magnetic bubble at $t$ = 3.5 ns (Figures 4a and 4b). Applying the electrical current continuously leads to the stabilization of the distorted magnetic bubble, eventually assuming a regular shape. At $t$ = 10 ns, a stable single magnetic bubble with vortex-like surface spin twists emerges, originating from the DDI (Figures 4a, 4b, and Supplemental Figure S4). Furthermore, the lower critical current density $j_{c1}$, corresponding to the transition process from the FM state to the magnetic bubble state, is independent of pulse width $w$, further confirming the electrical creation of a single magnetic bubble is driven by the STT effect (Figure 3c). Furthermore, due to the geometric constraint and the surface spin twists induced by DDI in 3D magnetic bubbles, the single magnetic bubble exhibits negligible motion when subjected to the applied pulsed current.

Notably, the uniform ferromagnetic (FM) state is unable to transition into magnetic bubbles under the influence of spin transfer torque (STT) because of its zero spatial magnetization gradient.[48] Actually, non-uniform magnetization caused by artificial defects or

WILEY-VCH

local impurities is considered the seed for creating skyrmions by STT.[48-52] The surface twists of the FM state in the nanostructured cell, induced by the magnetic dipole-dipole interaction, may aid in the formation of the magnetic bubble as seed domains.

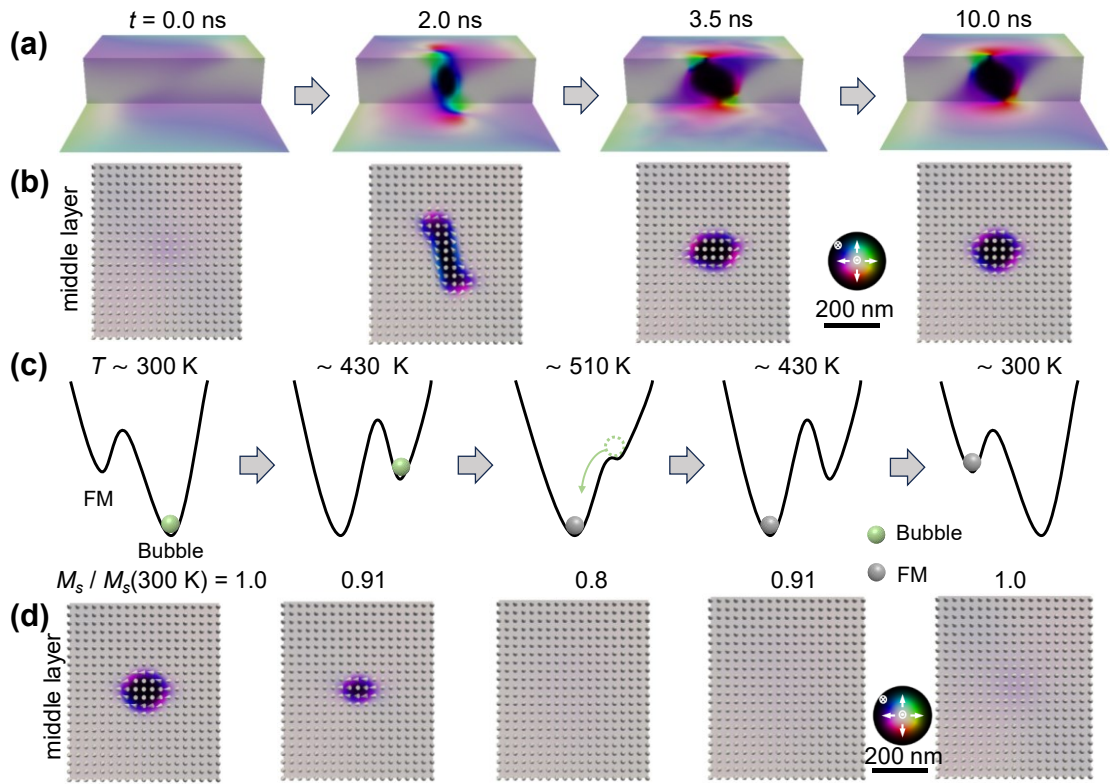

**Figure 4.** Simulated processes of writing and deleting a single magnetic bubble. a), b) Simulated magnetic structures in the cross-section and middle layer of the cell as a single bubble is created from the FM state by STT effect. Simulated current density $j = 50 \times 10^{10}$ A m$^{-2}$. c) Schematic diagram of energy density profiles as the temperature varies in a circular manner. d) Simulated magnetic structures in the middle layer of the cell corresponding to (c).

In Figures 3a and 3b, a single magnetic bubble annihilates, turning to a FM state after the application of a relatively high pulsed current. High current densities typically result in significant Joule heating, leading to a remarkable increase in the sample's temperature. The critical current density $j_{c2}$ required for the magnetic bubble to FM transition decreases as the pulse width increases, suggesting that the process is dominated by Joule heating effects. A longer pulse width necessitates a smaller current density to achieve equivalent Joule heating (Figure 3c). A previous study reported that a pulsed current with a current density $j$ of $1.7 \times 10^{11}$ A m$^{-2}$ and a pulse width $w$ of 20 ns will increase the temperature of the sample rapidly by about 350 K, consequently reducing the saturated magnetization $M_s$.[40, 53] Consequently, we conducted simulations to analyze the different magnetic spin textures resulting from Joule heating-induced variations in saturated magnetizations and presented schematic energy profiles (Figures 4c and 4d). Setting the single magnetic bubble as the initial state at $B$ = 365 mT, a 9% reduction in saturated magnetization results in the FM state having lower energy than the magnetic bubble state, indicating its greater stability (Figures 4c and 4d). When the $M_s$ decreases to 80% of its value at 300 K, the single magnetic bubble transforms into an FM state at a fixed field of $B$ = 365 mT (Figures 4c and 4d). After the pulsed current finishes, the sample's temperature returns to 300 K, and the FM state persists due to the absence of additional energy required to overcome the energy barrier between it and the magnetic bubble state (Figures 4c and 4d).

The critical current densities $j_{c1}$ and $j_{c2}$ are very sensitive to the magnetic field $B$. At $B$ = 543 mT, $j_{c1}$ and $j_{c2}$ are $2.66 \times 10^{10}$ A m$^{-2}$, $7.99 \times 10^{10}$ A m$^{-2}$, respectively (Figure 3d). As the magnetic field increases, $j_{c1}$ increases rapidly from $2.66 \times 10^{10}$ A m$^{-2}$ at $B$ = 543 mT to $6.47 \times 10^{10}$ A m$^{-2}$ at $B$ = 548 mT. When the magnetic field is above 548 mT, the FM state is unable to transform into a magnetic bubble by pulsed currents because $j_{c1}$ is possibly larger than $j_{c2}$ at that field, leading to the sufficient Joule heating that decreases the saturated magnetization and stabilizes the FM state. On the other hand, the stability of the FM state increases with the magnetic field for $j_{c2}$, requiring less reduced saturated magnetization and resulting in a decrease in $j_{c2}$.

**2.3 Deterministic Electrical Control of Single Magnetic Bubbles Repeatably**

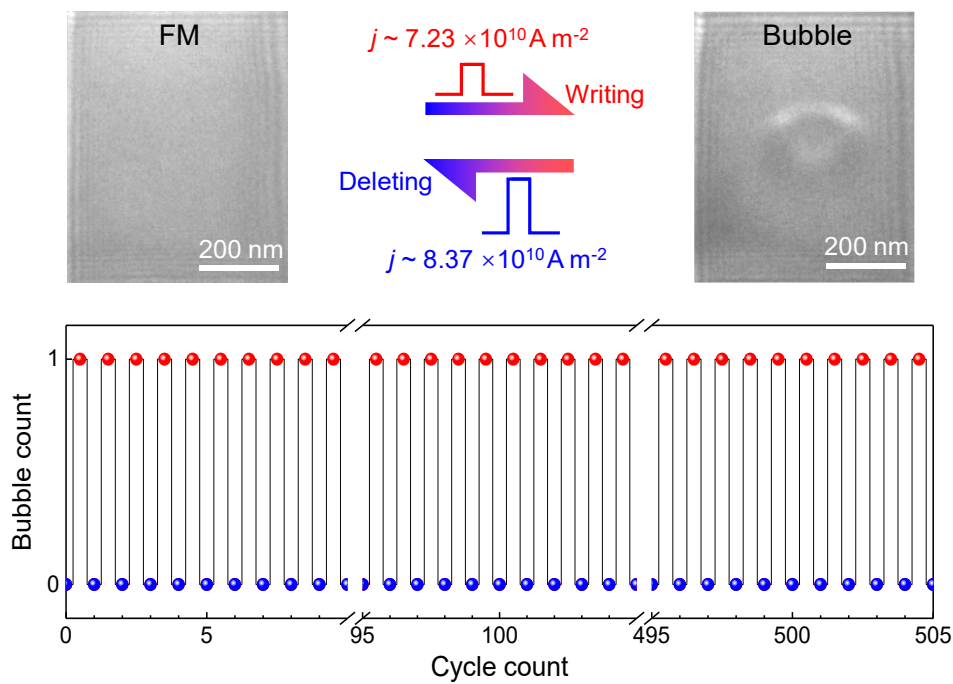

**Figure 5.** Repeatability test of electrical manipulation of a bubble. Two fixed current densities with an 80 ns pulse width are used to write and delete a single bubble. A lower current density, $7.23 \times 10^{10}$ A m$^{-2}$, corresponds to writing a bubble, represented by red circles in the diagram. A higher current density, $8.37 \times 10^{10}$ A m$^{-2}$, corresponds to deleting a bubble, represented by blue circles in the diagram. In each cycle, there is a time interval, about 1.5 s, between two adjacent pulsed currents.

According to the transition process between the FM state and the bubble state mentioned above, we select two discrete current densities to establish a writing-deleting cycle for the nanostructured cell. With a pulse width $w$ of 80 nanoseconds, a current density of $7.23 \times 10^{10}$ A m$^{-2}$ is chosen for writing a single magnetic bubble from the FM state, while a current density of $8.37 \times 10^{10}$ A m$^{-2}$ is selected for a reverse transition. By alternately applying these two pulsed currents to the sample, we are able to achieve a writing and deleting cycle of at least 500 repetitions without any errors (**Figure 5** and Supplemental Video S3). These findings suggest that the electrical manipulation of magnetic bubbles is a more viable approach compared to

WILEY-VCH

using oscillating magnetic fields or other techniques, offering a promising strategy for future bubble storage devices.

## 2.4 Detecting a Single Bubble through the AMR Effect

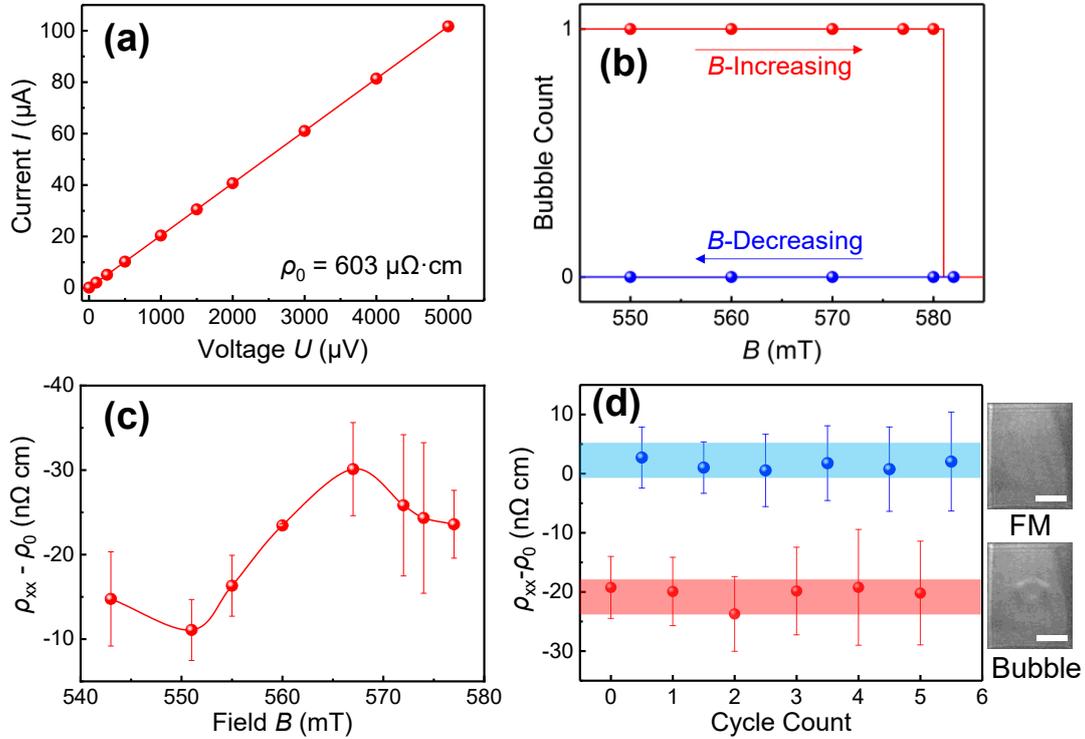

**Figure 6.** Detecting a single bubble from magnetoresistance measurement. a) The dependence of current $I$ on voltage $U$ in the FM state at a perpendicular field $B$ = 590 mT. $\rho_0$ is resistivity. b) The different bubble count at a fixed magnetic field $B$ = 577 mT. c) Experimental data of magnetoresistance changes with increasing magnetic fields. The titled angle between the normal direction of the sample and the magnetic field is ≈ 3°. d) The variation of magnetoresistance between FM state and bubble at a fixed field, $B$ = 577 mT. The red circles represent signals of the bubble and the blue circles represent the FM state. Each circle corresponds to a mean value of 50 consecutive measured points. Scale bars, 200 nm.

An in-situ electrical detection of a single magnetic bubble in a Lorentz TEM was performed using the standard four-probe method (Figure 1a). Setting the current at approximately 57 μA (current density $j \approx 5.35 \times 10^8$ A m$^{-2}$), the magnetoresistance (MR) was measured by reading out the voltage values from two central electrodes, in the hysteresis region of the magnetic phase diagram.

The flow of electrical current through a sample can result in Joule heating, potentially altering the sample's resistivity. Therefore, the *I-V* characteristic curve of the sample was initially measured to ensure the accuracy of the electrical detection (**Figure 6**a). As shown in Figure 6a, the voltage is correlated linearly with the current, indicating that the resistivity remains constant and suggesting that the current density utilized results in negligible Joule heating. The resistivity of the sample is approximately 603 μΩ cm, which is roughly twice as large as previously reported resistivity values for bulk $Fe_3Sn_2$.[42] This discrepancy may be attributed to the inhomogeneities and damage caused during the fabrication process using FIB.

In ferromagnets, the anisotropic magnetoresistance (AMR) effect takes a dominant role in MR measurement, which has been used to distinguish the existence of the magnetic spin textures indirectly.[54-58] It characterizes the longitudinal resistivity in dependence on the magnetization orientation relative to the current direction. The AMR effect is typically represented by the following equation[59, 60]:

$$\rho_{xx} = \rho_0 + A(\boldsymbol{n_I} \cdot \boldsymbol{n_m})^2 \tag{1}$$

Where $\rho_{xx}$ is the anisotropic magnetoresistance measured through two voltage probes, $\rho_0$ is the initial resistivity of the FM state at a saturated field, $\boldsymbol{n_I}$ is the unit vector along the current direction, $\boldsymbol{n_m}$ is the unit vector along the magnetization orientation, and *A* is the AMR coefficient. For instance, consider the case of the uniform FM state, where $\boldsymbol{n_m}$ aligns with the *z*-axis and $\boldsymbol{n_I}$ aligns with the *x*-axis, the AMR coefficient *A* can be determined through fitting the titled field direction relative to the magnetization orientation. Therefore, the parameter $\varphi$, defined as the angle between field *B* and the *c*-axis of the sample plane, is equal to the titled angle of the sample (Figure 1b), establishing the relationship between $\Delta\rho_{xx}$ and $\varphi$ through fitting with equation (1):

$$\Delta\rho_{xx} = \rho_{xx} - \rho_0 = A\left[\cos\left(\frac{\pi}{2} - \varphi\right)\right]^2 = A\sin(\varphi)^2 \qquad (2)$$

We set the magnetic field $B = 1750$ mT, significantly higher than the saturation field of the sample, to ensure proper alignment of magnetization with the magnetic field. Then the sample is tilted in the range of ≈ 50°, and the resistivity change is recorded simultaneously (Supplemental Figure S5). The data points fit a $\sin^2 x$ function well, indicating the changes of MR in the sample are mainly attributed to the AMR effect. The coefficient $A$, determined from the fitted curve, is inferred to be a negative value of $A = -9293$ n$\Omega$ cm, suggesting a decrease in resistivity as the sample is tilted.

Keeping the magnetic field $B = 577$ mT, both the single bubble state and ferromagnetic (FM) state can be achieved by adjusting the magnetic field within the hysteresis region (Figure 6b). For example, a single bubble can stably exist at 577 mT and transform into the FM state when the field exceeds 580 mT. Once the field is reduced back to 577 mT, the FM state remains stable. A noticeable difference in resistivity was observed between the bubble state and the FM state (Figure 6d). Each data point in the graph represents an average value of approximately 50 consecutive measurements of magnetoresistance (MR) signals, with error bars provided. Based on the six cycles of measurement conducted, it can be deduced that the MR signals for a single magnetic bubble exhibit a variation of approximately –22 n$\Omega$ cm. It is worth noting that the resistivity change caused by a single bubble yields a negative value, suggesting that in-plane magnetization has a slightly greater impact on electron scattering compared to the out-of-plane magnetization component. Moreover, the resistivity changes resulting from a single magnetic bubble were further examined across various magnetic field strengths within the hysteresis region. As the magnetic field strength increases, the resistivity change demonstrates a consistent upward trend, with the greatest observed value of approximately –30 nΩ cm occurring at 567 mT (Figure 6c).

## 3. Conclusions

In summary, we performed in-situ electrical manipulation of a single magnetic bubble in a nanostructured sample, achieving the functions of writing and deleting a single bubble via pulsed current with two different current densities. Applying these two current densities alternately, the writing and deleting processes can be realized at least 500 times with zero error rate, which may promote the bubble-based spintronics. We also detected the variation in MR of a single magnetic bubble, approximately –22 nΩ cm, with the measured resistivity change being primarily attributed to the anisotropic magnetoresistance (AMR) effect. Our findings offer a practical approach for writing, deleting, and detecting a bubble, enhancing the development of future nanostructured devices.

## 4. Experimental Section

*Sample Preparation*: Single crystals of $Fe_3Sn_2$ were grown using the chemical vapor transport method from a stoichiometric mixture of Fe (>99.9%) and Sn (>99.9%).[27] The 165-nm-thick nanostructured cell of $Fe_3Sn_2$, mounted on an in-situ electrical chip, was fabricated via a standard lift-out method from the bulk by using a focused ion beam instrument (Helios Nanolab 600i, FEI).

*TEM Measurements*: Fresnel magnetic images were recorded by a TEM instrument (Talos F200X, FEI) operating at 200 kV in Lorentz mode. The perpendicular magnetic field was adjusted by manipulating the objective current, with pulsed currents being supplied by a voltage source (AVR-E3-B-PN-AC22, Avtech Electrosystems). The resistance of the cell was measured using the standard lock-in technique (SR830) to minimize measurement noise. All experiments were conducted at room temperature.

*Micromagnetic Simulations*: Micromagnetic Simulations were performed by using a GPU-

accelerated program, Mumax3.[61] The exchange interactions, uniaxial magnetic anisotropy, Zeeman energy, and dipole-dipole interactions were considered in the simulations. Magnetic parameters were set based on the $Fe_3Sn_2$ material with $A = 8.25$ pJ m$^{-1}$, $K_u = 54.5$ kJ m$^{-3}$, and saturation magnetization $M_s = 622.7$ kA m$^{-1}$.[27] The length, width, and thickness of simulated geometry were set as 550, 650, and 150 nm, respectively. The cell size was set to $5 \times 5 \times 5$ nm$^3$. The equilibrium spin configurations were obtained using the conjugate gradient method.

**Supporting Information**

Supporting Information is available from the Wiley Online Library or from the author.

**Acknowledgments**

This work was supported by the National Key R&D Program of China, Grant No. 2022YFA1403603; the Natural Science Foundation of China, Grants No. 52325105, 12422403, 12174396, 12104123, 12241406, and U24A6001; the Anhui Provincial Natural Science Foundation, Grant No. 2308085Y32 and 2408085QA022; the Natural Science Project of Colleges and Universities in Anhui Province, Grants No. 2022AH030011 and 2024AH030046; CAS Project for Young Scientists in Basic Research, Grant No. YSBR-084; Systematic Fundamental Research Program Leveraging Major Scientific and Technological Infrastructure, Chinese Academy of Sciences, Grant No. JZHKYPT-2021-08; Anhui Province Excellent Young Teacher Training Project, Grant No. YQZD2023067; the 2024 Project of GDRCYY (No. 217, Yaodong Wu); and the China Postdoctoral Science Foundation, Grants No. 2023M743543 and 2024M760006.

**Author Contributions**

H. D. and J. T. supervised the project. J. T. conceived the idea and designed the experiments. Y.

Wang, J. L., and Y. X. synthesized the $Fe_3Sn_2$ single crystals. J. J. fabricated the $Fe_3Sn_2$ nanostructured cells. J. J., J. T., Y. Z., S. Q., H. Z., and Y. Wu performed the TEM measurements. J. T. and Y. Wu performed the simulations. J. T., J. J., Y. Wu, and H. D. wrote the manuscript with input from all authors. All authors discussed the results and contributed to the manuscript.

**Conflict of Interest:**

The authors declare no competing financial interest.